# CONTEMPORARY SEMANTIC WEB SERVICE FRAMEWORKS: AN OVERVIEW AND COMPARISONS

Keyvan Mohebbi[1], Suhaimi Ibrahim[2], Norbik Bashah Idris[3]

[1]Faculty of Computer Science and Information Systems,
Universiti Teknologi Malaysia (UTM), Malaysia
mkeyvan2@live.utm.my
[2,3]Advanced Informatics School (AIS), Universiti Teknologi Malaysia (UTM), Malaysia

## ABSTRACT

*The growing proliferation of distributed information systems, allows organizations to offer their business processes to a worldwide audience through Web services. Semantic Web services have emerged as a means to achieve the vision of automatic discovery, selection, composition, and invocation of Web services by encoding the specifications of these software components in an unambiguous and machine-interpretable form. Several frameworks have been devised as enabling technologies for Semantic Web services. In this paper, we survey the prominent Semantic Web service frameworks. In addition, a set of criteria is identified and the discussed frameworks are evaluated and compared with respect to these criteria. Knowing the strengths and weaknesses of the Semantic Web service frameworks can help researchers to utilize the most appropriate one according to their needs.*

## 1. INTRODUCTION

The Service Oriented Architecture (SOA) is considered to be the latest development of a long series of advancements in software engineering addressing the reuse of software components [1]. Web services can be seen as the technical solutions to implement the SOA vision. A Web service as defined by the World Wide Web Consortium (W3C) is a " software system identified by a URI [2], whose public interfaces and bindings are defined and described using XML. Its definition can be discovered by other software systems. These systems may then interact with the Web service in a manner prescribed by its definition, using XML based messages conveyed by Internet protocols" [3].

The current Web lacks a proper support to its users when it comes to finding, extracting, and combining information. The main reason is that the meaning of Web content is not machine-accessible. The Semantic Web extends the current Web providing machine-processable semantics to Web resources. The backbone of Semantic Web are ontologies [4]. They provide greater expressiveness when modeling domain knowledge, thus facilitate knowledge reuse and sharing between heterogeneous and distributed applications.

Semantic Web services aim at making Web services machine-processable and intelligent. This can be realized using Semantic Web technologies for Web service annotation and processing. The idea is to provide ontology-based descriptions of Web services that could be processed by ontology reasoning tools. In that way, intelligent agents would be able to automatically understand what a Web service does and what it needs in order to perform a task [5]. The most common usage tasks during the Web services' life cycle include their discovery, selection, composition, dynamic binding, and invocation. Adding semantics to Web services is also considered as a major factor which facilitates their foundations, managements, and engineering [6].

There are two ways of creating semantic annotated Web services. One way is to create an independent semantic Web service description framework and link it to the current Web services standards. The leading research efforts are OWL-S [7] and WSMO [8]. The other way is to add semantic annotations into the current Web services standards. The major research works in this way are WSDL-S [9] and SAWSDL [10].

Generally, Web services are described using the Web Service Description Language (WSDL) [11]. WSDL is an XML based, low level syntactical, and developer oriented service description language. The main idea of the existing frameworks is to build a semantic layer either on the top of WSDL or to be integrated into WSDL to semantically describe the capabilities of Web services so that a software agent or other services can find out about a Web service's capabilities and how it can be used.

In this paper, we survey the prominent frameworks for enabling Semantic Web services. Further, we introduce and describe a set of several criteria. Then, the reviewed frameworks are evaluated and compared according to those criteria.

The results of this study can help researchers in the field to select the most appropriate Semantic Web service framework matching to their needs. Such a primary decision would affect the designation or implementation of each of the Web services' usage tasks and should be made prior to them.

The rest of this paper is structured as follows. We review related works on the comparison of Semantic Web service frameworks in section 2. Then, we provide an overview on prominent Semantic Web service frameworks in sections 3 to 6. After introducing some criteria and describing them, we evaluate and compare the discussed frameworks in section 7. Our conclusion is presented in section 8.

## 2. RELATED WORKS

To the best of our knowledge, a very few number of research works have been published so far concerning the evaluation and comparison of Semantic Web service frameworks.

In [11], a conceptual comparison is conducted that identifies the overlaps and differences of OWL-S and WSMO. Although a wide range of aspects has been proposed for such comparison, but it does not include WSDL-S and SAWSDL frameworks. Cabral et al. [6], compare three enabling technologies for Semantic Web services, namely IRS-II, OWL-S, and WSMF. They characterize the infrastructure of Semantic Web services along three orthogonal dimensions: activities, architecture, and service ontology. Their comparison then focuses on assessing the delivered components of three considered approaches with respect to the mentioned dimensions. One of the most comprehensive surveys on Semantic Web service frameworks is provided in [12]. It presents an overview and a comparative analysis of OWL-S, WSMO, and WSDL-S framework. However, it does not consider SAWSDL. In [13] a comparison of OWL-S, WSMO, and SAWSDL frameworks is presented which points to the syntactic and semantic heterogeneities of these frameworks. However, the authors neither describe the used criteria, nor provide an overall result.

Our work is mostly inspired by [11] and [12]. We summarize eight criteria from the past literature and analyze four contemporary yet prominent Semantic Web service frameworks in terms of these criteria. Such comparison is not addressed by any of the aforementioned works.

## 3. OWL-S (ONTOLOGY WEB LANGUAGE FOR SERVICES)

OWL-S [7] defines an upper ontology for describing the properties and capabilities of Web services in OWL in order to facilitating the automation of Web service tasks, including Web service discovery, execution, composition and interoperation. According to IEEE P1600.1 [14]: "An upper ontology is limited to concepts that are meta, generic, abstract and philosophical, and therefore are general enough to address (at a high level) a broad range of domain areas. Concepts specific to given domains will not be included; however, this standard will provide a structure and a set of general concepts upon which domain ontologies (e.g. medical, financial, engineering, etc.) could be constructed."

The structure of the upper ontology of services is divided into three main parts, as shown in Figure 1.

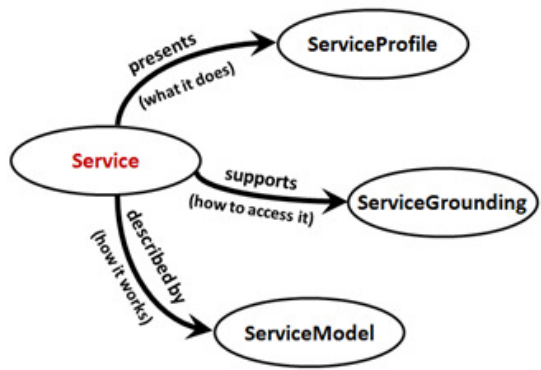

Figure 1. Service Ontology in OWL-S

The class Service provides an organizational point of the reference for a declared Web service; one instance of Service will exist for each distinct published service. The classes *ServiceProfile*, *ServiceModel*, and *ServiceGrounding* are the respective ranges of the properties presents, describedBy, and supports of a Service. Each of these perspectives provides an essential type of information to know about the service, as explained below:

*ServiceProfile*: tells what the service does, in a way that is suitable for a service-seeking agent (or matchmaking agent acting on behalf of a service-seeking agent) to determine whether the service meets its needs, thus enabling discovery and matchmaking. The profile comprises functional and nonfunctional aspects of the service. The functional description of the service is expressed in terms of the transformation produced by the service. It includes information transformation represented by inputs and outputs and the change in the state of the real world caused by the execution of the service which is represented by preconditions and effects. Non-functional properties include references to existing categorization schemes (e.g. UNSPC) or ontologies, provider information, the quality rating of the service and so on.

*ServiceModel*: tells how a service works. This model is used to enable service invocation, enactment, composition, monitoring, and recovery. The service model views the interactions of the service as a process. A process is not necessarily a program to be executed, but rather a specification of ways in which a client may interact with a service. OWL-S distinguishes between two categories of services, namely atomic and composite. Atomic services are ones where a single Web-accessible computer program, sensor, or device is invoked by a request message, performs its task and perhaps produces a single response to the requester. With atomic services there is no ongoing interaction between the user and the service. In contrast, complex or composite services are built up from multiple more primitive services by workflow structures to determine the control flow, and may require an additional interaction or conversation between the requester and the set of services that are being utilized.

*ServiceGrounding*: specifies the details of how an agent can access a service. Typically a grounding maps the constructs of the process model to detailed specifications of message formats, protocols, and so forth. In detail, OWL-S allows one to map atomic processes to WSDL operations and their inputs and outputs to WSDL messages. Although WSDL is the only completely defined grounding for OWL, OWL-S does not restrict itself to WSDL as the only underlying service technology, rather it is extensible to other grounding mechanisms.

In general, the *ServiceProfile* provides the information needed for an agent to discover a service, while the *ServiceModel* and *ServiceGrounding*, taken together, provide enough information for an agent to make use of a service.

## 4. WSMO (WEB SERVICE MODELING ONTOLOGY)

Web service Modeling Ontology (WSMO) [8] provides a conceptual framework and a formal language for semantically describing Web services in order to facilitate the automation of discovering, combining and invoking electronic services over the Web. WSMO is a meta-model for Semantic Web services related aspects. The Meta-Object Facility (MOF) [15] specification is used as a basis to this model. MOF defines an abstract language and framework for specifying, constructing, and managing technology neutral meta-models.

WSMO takes the Web Service Modeling Framework (WSMF) [16] as basis and further refines and extends its concepts through a formal ontology and language. The Web service Modeling Language (WSML) [17] is a family of languages which formalizes WSMO.

Web Service Modeling Execution Environment (WSMX) [18] has developed to provide a reference architecture and implementation for the dynamic discovery, selection, mediation, invocation and inter-operation of Semantic Web services based on the WSMO specification.
Following the main elements identified in the WSMF, WSMO distinguishes four top level elements as the main concepts which have to be described in order to explain Semantic Web services [8]. These elements are depicted in Figure 2 and defined in the following.

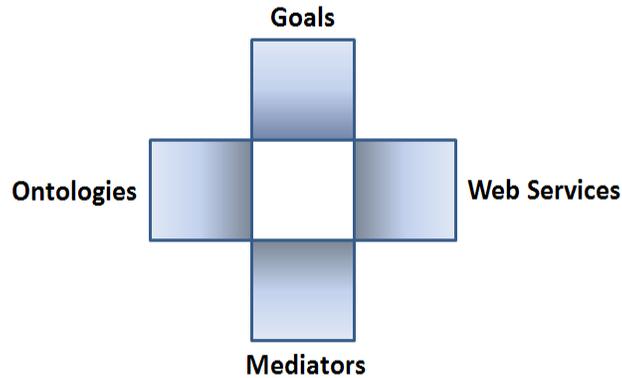

Figure 2. The Top-Level Elements of WSMO

*Ontologies*: define a shared common terminology by providing concepts, and relationships between the concepts. In order to capture semantic properties of relations and concepts, an ontology also provides a set of axioms, which are expressions in some logical language. In WSMO, ontologies provide the terminology used by other elements to describe the relevant aspects of the domains of discourse.

*Web services*: describe the computational entity providing access to services. A service is the actual value in a domain. Web service descriptions consist of their capabilities, interfaces and internal working. All these aspects of a Web service are described using the terminology defined by the ontologies. In order to allow their discovery, invocation, composition, execution, monitoring, mediation, and compensation, Web services are described in WSMO from three different points of view: non-functional properties, functionality and behavior. The capability of a Web service encapsulates its functionality and an interface of a Web service describes its behavior in terms of communication and collaboration.

*Goals*: describe user desire aspects with respect to the requested functionality as opposed to the provided functionality described in the Web service capability.

*Mediators*: describe elements that aim to overcome interoperability problems that appear between different WSMO elements. Mediators are the means to resolve incompatibilities that arise on the data, process and protocol level. Mismatches between different used terminologies are on data level, in Web services (and Goals) combinations are on process level and in Web services communications are on protocol level.

## 5. WSDL-S (WEB SERVICE DESCRIPTION LANGUAGE-SEMANTICS)

Building on the descriptive capability of WSDL [19], WSDL-S [9] is a lightweight approach to enrich Web services with semantics. In WSDL-S, the semantic models are maintained outside of WSDL documents and are referenced from the WSDL document via WSDL extensibility elements.

WSDL-S provides mechanisms to annotate the service and its inputs, outputs and operations. Additionally, it provides mechanisms to specify and annotate preconditions and effects of Web services. These elements allow one to define the conditions that must hold before executing an operation and the effects the execution would have. The preconditions and effects together with the semantic annotations of inputs and outputs can be used to automate the process of service discovery. Moreover, in order to define categorization information for publishing Web services in registries such as UDDI, WSDL-S considers the use of category extension attribute on the interface element [9].

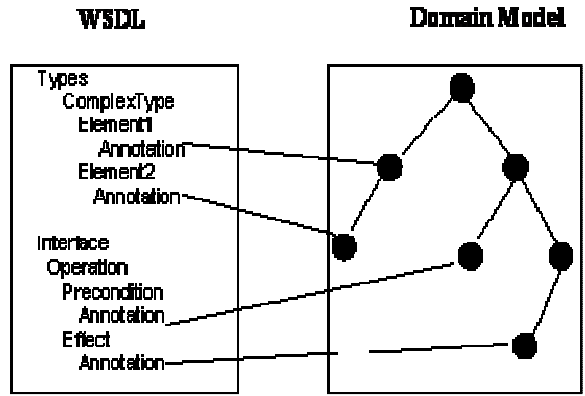

Figure 3. Externalized Representation and Association of Semantics to WSDL Elements

Figure 3 captures the essence of WSDL-S. It highlights how the domain semantic model which may consists of one or more ontologies, is kept external to a Web service model and how semantic annotations are associated with various elements of a WSDL document (including inputs, outputs and functional aspects like operations, preconditions and effects).

The key design principles for WSDL-S are as follows [9]:

- The Web services standards are fast becoming a preferred technology for application integration because of the promise of their interoperability. The authors of WSDL-S aim to adding semantics to Web services in an upwardly compatible manner based on existing Web services standards without disrupting the existing install-base of Web services.

- WSDL-S is independent of the language used for defining the semantic models. By keeping the semantic annotation mechanism separate from the representation of the semantic descriptions, the approach supports user's choice of any semantic representation language.

- WSDL-S allows multiple annotations written in different semantic representation languages to be associated with Web services. Annotating a service in multiple semantic representation languages will lead it to be discovered by multiple discovery engines.

- A common practice in the integration of Web services is to reuse interfaces that are described in XML. XML schema-based types are widely used in the definition of business documents. Supporting semantic annotations of Web services whose data types are described in XML, WSDL-S allows the gradual upgrade of deployed WSDL documents to include semantics.

## 6. SAWSDL (SEMANTIC ANNOTATIONS FOR WSDL)

Semantic Annotations for WSDL and XML Schema SAWSDL [10] defines mechanisms for adding semantic annotations to WSDL components. SAWSDL is not a language for representing the semantic models rather it provides mechanisms by which WSDL components can reference concepts from the semantic models that are defined either within or outside the WSDL document [20]. SAWSDL is a restricted version of on an earlier proposal WSDL-S, submitted to W3C in which annotations like preconditions and effects have not been explicitly considered. However, SAWSDL does not prevent the use of these annotations.

The key design principles for SAWSDL are as follows:

- The specification enables semantic annotations for Web Services based on the extensibility framework of WSDL.

- It is agnostic to both semantic representation and mapping languages.
- It enables semantic annotations for Web Services both for discovering and for invoking Web Services.

- SAWSDL annotates WSDL interfaces and operations with categorization information that can be used to publish a Web service in a registry. The annotations on schema types can be used during Web service discovery and composition. In addition, SAWSDL defines an annotation mechanism for specifying the data mapping of XML Schema types to and from an ontology; such mappings could be used during invocation. The annotation mechanism is independent of the ontology expression language and requires no particular of such languages. It is also independent of mapping languages and does not restrict the possible choices of mapping languages [20].

- The disadvantage of existing Semantic Web service formalisms is that they take a top down approach to model Web services concepts without considering WSDL. SAWSDL takes a bottom-up approach building on top of WSDL. Because SAWSDL is agnostic to ontology language used to model Web service aspects, it is possible to use different Semantic Web service formalisms according to the needs of a particular domain.

- SAWSDL specification addresses the problem of adding data semantics by annotating WSDL input/output messages and XML schema types. Moreover, SAWSDL adds functional semantics through supporting the categorization of the Web service interface. But still SAWSDL does not model other functional semantics e.g. pre and post-conditions of a service [21].

## 7. THE COMPARISON OF SEMANTIC WEB SERVICE FRAMEWORKS

In this section, first we define some criteria and then evaluate and compare the aforementioned Semantic Web service frameworks according to these criteria. The result can be seen in Table 1.

Table 1. Comparison of Semantic Web service Frameworks

| Criteria | Semantic Web Service Framework | | | |
|---|---|---|---|---|
| | **OWL-S** | **WSMO** | **WSDL-S** | **SAWSDL** |
| Supported Elements for Annotating Semantically | Service, Operation, Input/Output, Pre/Post-Condition | Service, Operation, Input/Output, Pre/Post-Condition | Service, Operation, Input/Output, Pre/Post-Condition | Service, Operation, Input/Output |
| Directly Supporting Non-Functional Properties | Service Profile | All Elements | No Element | No Element |
| Relation with WSDL | Defines connectivity to WSDL via Grounding Model | Defines connectivity to WSDL via Grounding Model | Specifies annotations directly in WSDL | Specifies annotations directly in WSDL |
| Support Complex Services (Processes) | Yes | Yes | No | No |
| Ontology Description Language | OWL | WSML | User Choice | User Choice |
| Formalism | DL | DL-FOL-LP | User Choice | User Choice |
| Support Multiple Semantic Annotations for Services | No | No | Yes | Yes |
| Support Conditional Result | Yes | Yes | No | No |
| Overall Result | **Good** | **Good** | **Medium** | **Medium** |

## 7.1. Definition of Comparative Evaluation Criteria

*1) Supported Elements for Annotating Semantically*

Indicates the elements that are considered to annotate by the framework. This will affect service discovery as the more elements annotates semantically, the more precise discovery will be.

*2) Directly Supporting Non-Functional Properties*

Indicates whether the framework supports expressing non-functional properties or not. Taking into account non-functional properties in addition to functional properties (inputs, outputs, pre-conditions, post-conditions), leads to a more precise service discovery.

*3) Relation with WSDL*

Indicates the way that a framework semantically extends traditional Web service descriptions in WSDL as an important industry standard.

*4) Support Complex Services (Processes)*

Indicates whether the framework supports complex services in addition to default atomic ones. An atomic service is that invoked by a requester and perhaps returns a single response, with no ongoing interaction between the requester and the service. Complex service is composed of multiple primitive services, and may require a set of interactions between the requester and the service.

*5) Ontology Description Language*

Indicates the ontology description language used by the framework.

*6) Formalism*

Indicates the formal logic language used by the framework.

*7) Support Multiple Semantic Annotations for Services*

Indicates whether the framework supports multiple semantic annotations for services or not. Using this feature, providers may choose to annotate their services in multiple semantic representation languages to be discovered by multiple discovery engines [9].

*8) Support Conditional Result*

Indicates whether the framework supports conditional result for operations or not. Using this feature the framework provides better representation in terms of outcome or effect.

## 7.2. Discussion

If a framework does address majority of the criteria or fulfill them in a more complete manner, has been evaluated as "*Good*" and if it addresses some of the criteria or fulfill them in a less complete manner, has been evaluated as "*Medium*". This way, OWL-S and WSMO are evaluated as "*Good*" frameworks, while WSDL-S and SAWSDL are evaluated as "*Medium*". Evaluating of these frameworks with respect to each criterion is explained as follows:

*1) Supported Elements for Annotating Semantically*

All of the frameworks support semantically annotating Services, Operations, Inputs and Outputs. However, unlike OWL-S, WSMO and WSDL-S, SAWSDL does not support the definition of service's Pre- and Post-conditions. This seriously limits the annotation possibilities for the functional behavior of Web services. For a more precise service discovery, Pre- and Post-conditions are essential [22].

*2) Directly Supporting Non-Functional Properties*

Non-functional properties in OWL-S are restricted to the Service Profile. However, these can be expressed in any WSMO element. On the other hand, WSDL-S and SAWSDL do not support such properties directly rather points to several existing proposals for standards from the Web service community [23].

*3) Relation with WSDL*

While OWL-S and WSMO define necessary links to WSDL via their Grounding Model to use its invocation model, WSDL-S and SAWSDL specify semantic annotations directly in WSDL as extensibility elements [24]. Compared with other two, WSDL-S and SAWSDL do not claim to be a fully-fledged description framework/ ontology rather they are minimalist approaches which aim at a direct extension of the existing "traditional" Web service descriptions in WSDL with semantics. Extending the industry standards such as WSDL to include semantics is a more practical approach for adoption [23].

*4) Support Complex Services (Processes)*

OWL-S presents a framework for simple (atomic) as well as complex (process) services. WSMO includes frameworks for Web service choreography and orchestration. WSDL-S and SAWSDL exclude the process specification models (i.e. complex services) from the scope [24].

*5) Ontology Description Language*

By keeping semantic model outside WSDL, WSDL-S and SAWSDL are impartial in relation to any ontology representation language. The advantage of this approach is that it builds upon and stays close to the existing industry standards. But on the other hand, without a certain degree of commitment to a specific language, or a definition of how different semantics of usable languages relate to one another, it is impossible to formally define requests, queries, or notions of a "match" between service requests and service descriptions [23].

*6) Formalism*

For formal logic language, OWL-S relies on Description Logics and WSMO relies on Description Logics, First-Order Logic and Logic Programming. On the other hand, both WSDL-S and SAWSDL are agnostic to ontology language. Therefore, users can freely choose any formalism [24].

*7) Support Multiple Semantic Annotations for Services*

Only WSDL-S and SAWSDL allow multiple annotations to be associated with Web services.

*8) Support Conditional Result*

OWL-S and WSMO use a conditional result mechanism and hence, operations may result in different outcomes. OWL-S' conditional outputs and effects can be expressed in WSMO, as post-conditions and effects are defined using a logical expression. However, WSDL-S and SAWSDL lack the notion of conditional outputs and effects [9].

## 8. CONCLUSIONS

In this paper, we aimed to provide an overview and compare the prominent Semantic Web service frameworks. All of these contemporary frameworks, namely OWL-S, WSMO, WSDL-S, and SAWSDL have been submitted to W3C as alternate proposals . After introducing a set of criteria, the reviewed frameworks are evaluated and compared according to those criteria.

Generally, there are two different paths to define a framework for enabling Semantic Web services. One path takes a revolutionary approach in that all aspects of semantic services are reconsidered. OWL-S and WSMO follow this line. The other path represents a more evolutionary approach that stays consistent with existing standards and industrial practices. Concrete examples are WSDL-S and SAWSDL.

In addition, the results of this study show that OWL-S and WSMO provide much richer semantic formalisms for Web services. In contrast, WSDL-S and SAWSDL provide quite a lightweight approach for such purpose. However, it is worth noting that SAWSDL is currently the only official standard for Semantic Web services recommended by W3C.

All of these Semantic Web services research efforts try to overcome the drawback of lack of semantics of the current Web services standards and realize automatic Web service discovery, composition, and invocation by providing appropriate description means that enable the effective exploitation of semantic annotations. However, semantically annotating functional components, i.e. Web services, is much more complicated than annotating static web information [25]. Therefore, further research efforts are still required in this area.

## ACKNOWLEDGEMENTS

This research is supported by Universiti Teknologi Malaysia (UTM) under the Vot. 00H74. The authors would like to thank UTM and Ministry of Higher Education (MOHE) Malaysia.